\documentclass[reprint,showpacs,amsmath,amssymb,aps]{revtex4}
\usepackage{graphicx,color}
\newcommand{\Array}[2]{\left(\begin{array}{#1}#2\end{array}\right)}

\begin{document}

\title{A Unified Yukawa Interaction for the Standard Model of Quarks and Leptons}
 \author{Ying Zhang \footnote{E-mail: hepzhy@mail.xjtu.edu.cn.}}
\address{Institute of Theoretical Physics, School of Physics, Xi'an Jiaotong University, Xi'an, 710049, China}
\date{\today}

\begin{abstract}
To address fermion mass hierarchy and flavor mixings in the quark and lepton sectors, a minimal flavor structure without any redundant parameters beyond phenomenological observables is proposed via decomposition of the Standard Model Yukawa mass matrix into a bi-unitary form. After reviewing the roles and parameterization of the factorized matrix ${\bf M}_0^f$ and ${\bf F}_L^f$ in fermion masses and mixings, we generalize the mechanism to up- and down-type fermions to unify them into a universal quark/lepton Yukawa interaction. 
In the same way, a unified form of the description of the quark and lepton Yukawa interactions is also proposed, which shows a similar picture as the unification of gauge interactions.
\end{abstract}

\maketitle

%%%%%%%%%%%%%%%%%%%
% flavor and flavor
%
%
%
%
%
%%%%%%%

\section{Motivation}
Although the Standard Model (SM) has been shown to be successful in phenomenology, some puzzles remain mysterious, especially in flavor physics, including the origin of CP violations, fermion mass hierarchy, flavor mixings, etc. \cite{Buras2005arXiv,Raidal2008EPJC,ZZX2020PR}. To address these interesting issues, the SM must be updated. Some creative motivations for this research are based on assuming mechanisms and models \cite{Feruglio2015EPJC,Rodejohann2003PRD,Li2005PRD}, but much are inspired by discrete symmetries and/or patterns \cite{King2013RPP}. Behind the new physics, different numbers of new particles and/or interactions are assumed. Unfortunately, to date, experiments have yielded null results in finding relevant information \cite{PDG2020}. This status hints at another way to improve the SM: re-construction using only currently found particles and interactions. Some clues can be extracted from the existing SM. Comparing the three kinds of fundamental gauge interactions with simple gauge couplings, Yukawa interactions include too many redundant parameters in Yukawa couplings. In fact, to respond to fermion masses and flavor mixings with CP violation (CPV) in both CKM and PMNS, Yukawa coupling ${\bf y}^f_{ij}$ is assigned a complex value. In the quark sector, for example, ${\bf y}^u_{ij}$ and ${\bf y}^d_{ij}$ include $(9+9)\times 2$ degrees of freedom (dof). However, in phenomenology, only 10 quantities are observable: 6 quark masses, 3 CKM mixing angles and 1 CP-violating phase. These observables are often regard as the SM input, whereas Yukawa couplings only appear in the building process of the  Lagrangian and retain unknown values. 
Another clue comes from the similarity between the quark and lepton sectors. Considering the minimal extended SM with the Dirac neutrino mass, the flavor characteristics in the quark sector are completely duplicated in the lepton sector. 
This encourages us to seek a common mechanism that underlies the quark and lepton flavors.
As the main characteristics, both hierarchal mass and flavor mixing with CPV point to the Yukawa term. In this paper, we start with the decomposition of the fermion mass matrix. A close-to-flat matrix is proposed to address hierarchal mass and guides unification of Yukawa couplings into a universal form for all SM fermions step by step. 

In the next section, we briefly review the minimal flavor structure with no redundant parameters exceeding the phenomenological observables proposed in \cite{Zhang2021arXiv}, which has successfully reproduced not only quark  masses and CKM but also lepton masses and PMNS. 
After bi-unitary decomposition of the fermion mass matrix, the real matrix ${\bf M}_0^f$ and unitary matrix ${\bf F}_{L,R}^f$ are responsible for the physical mass and the origin of CPV, respectively. 
This result inspires to parameterize ${\bf M}_0^f$ into a close-to-flat real symmetry matrix. The unitary matrix ${\bf F}_{L,R}^f$ can be absorbed into the fermion redefinition, which shows an origin of CPV from a quantum superposition phase between gauge eigenstates and Yukawa interaction states. In the Yukawa interaction states, family-universal couplings can arise as a natural result.
In Sec.\ref{sec.quarkleptonYuk}, we apply the same mechanism to scalar and right-handed quarks and find a similar flat matrix. The Yukawa couplings from up- and down-type fermions can be unified into a universal form for quarks (lepton) when a slight breaking effect is introduced. Continuing the same process in Sec.\ref{sec.unifiedYuk}, the quark and lepton Yukawa interactions are eventually unified into a common real coupling in appropriate Yukawa eigenstates.  As an example, we focus on the flavor structure of quarks. All formulae can be directly generalized to the lepton sector with minimal extended Dirac neutrinos (normal order). Finally, a summary is given.

%%%%
\section{Minimal Flavor Structure in a Family}\label{sec.MFS}

In ref. \cite{Zhang2021arXiv}, a family-universal Yukawa coupling is proposed by assuming flavor vacuum perturbations. The mechanism  has successfully been used to address not only  hierarchal masses of quarks and leptons but also CKM and PMNS mixings. Here, a short review is given, which will be used as a starting point to realize a universal Yukawa interaction for all SM fermions. 

The quark mass matrix in the SM is expressed as
\begin{eqnarray}
	{\bf M}_{ij}^q=\frac{v_0}{\sqrt{2}}{\bf y}_{ij}^q
\end{eqnarray}
with $q=u,d$ for up- and down-type quarks and Higgs vacuum expectation value $\langle H\rangle=v_0/\sqrt{2}$ .  As a complex matrix, ${\bf M}^q$ can be diagonalized  by  bi-linear unitary transformations
$\psi_{L,R}^q=({\bf U}^q_{L,R})^\dag \psi^{q,m}_{L,R}$ into mass eigenstates 
\begin{eqnarray}
{\bf M}^q\rightarrow {\bf U}_L^q{{\bf M}^q}({\bf U}_R^q)^\dag=\text{diag}(m^q_1,m^q_2,m^q_3)
\label{eq.diagonalquarkmass}
\end{eqnarray}
The CKM mixing matrix can be expressed as ${\bf U}_{CKM}={\bf U}^u_L{{\bf U}^d_L}^\dag.$
Thus, the quark flavor structure is completely encoded into a mass matrix generated from complex Yukawa couplings ${\bf y}^q_{ij}$.

As hinted  by  eq.(\ref{eq.diagonalquarkmass}), ${\bf M}^q$ can be decomposed into a bi-unitary form,
\begin{eqnarray} 
	{\bf M}^q={{\bf F}_L^q}^\dag {\bf M}_0^q {\bf F}^q_R
\label{eq.massmatrixdecomp}
\end{eqnarray}
with real matrix ${\bf M}_0^q$ and left- and right-handed unitary matrices ${\bf F}_{L,R}^q$, which relate to ${\bf U}_{L,R}^q$ by an $SO(3)$ rotation ${\bf U}_0^q$,
\begin{eqnarray}
	{\bf M}_0^q={\bf U}_0^q\text{diag}(m_1^q,m_2^q,m_3^q)({\bf U}_0^q)^T,~~
	{\bf U}_{L,R}^q={\bf U}_0^q{\bf F}_{L,R}^q
\end{eqnarray}
These factorized matrices play different roles in the quark mass and mixing:
\begin{itemize}
	\item Quark physical masses are completely determined by the eigenvalues of ${\bf M}_0^q$; neither the left- nor right-handed unitary transformation provides any contributions.
	\item Left-handed unitary matrix ${\bf F}^q_L$ provides complex phases required by the non-vanishing CP violation in CKM.
	\begin{eqnarray}
	{\bf U}_{CKM}&=&{\bf U}_0^u{{\bf {F}}_L^u}({\bf {F}}_L^d)^\dag({\bf U}_0^d)^\dag
	\label{eq.Uckm}
	\end{eqnarray}
	\item Right-handed unitary matrix ${\bf F}^q_R$  has no effect on both the quark mass and mixings (taken as ${\bf F}^q_R=1)$.
\end{itemize}
To address hierarchal mass eigenvalues, a close-to-flat flavor vacuum is assumed,
\begin{eqnarray}
		{\bf M}_0^f&=&\frac{m_{\Sigma}^f}{3}{\bf I}_\Delta^f
	,~~~
		{\bf I}_\Delta^f=\left(\begin{array}{ccc}1 & 1+\delta_{12}^f & 1+\delta_{13}^f\\ 1+\delta_{12}^f & 1 & 1+\delta_{23}^f \\ 1+\delta_{13}^f & 1+\delta_{23}^f & 1\end{array}\right)
	\label{eq.Mpara}
	\end{eqnarray}
where $\delta^f_{ij}$ represents a slight perturbation. A possible origin of non-diagonal perturbations $\delta_{ij}^f$ may be vacuum corrections for different flavor Yukawa interaction.
%%%
The physical masses are controlled by $\delta_{ij}^f$,
\begin{eqnarray}
		m^f_{1,2}&=&\frac{y^fv_0}{\sqrt{2}}\left(\frac{1}{3}S^f\mp\frac{2}{3}\sqrt{Q^f}\right)+\mathcal{O}(\delta^2)
		\label{eq.m1m2}\\
		m^f_3&=&\frac{y^fv_0}{\sqrt{2}}\left(3-\frac{2}{3}S^f\right)+\mathcal{O}(\delta^2)
		 \label{eq.m3}
	\end{eqnarray}
with parameters $S^f$ and $Q^f$ given as
	\begin{eqnarray}
		S^f&\equiv& -\delta^f_{12}-\delta^f_{23}-\delta^f_{13}
\label{eq.S}\\
		Q^f&\equiv&(\delta^f_{12})^2+(\delta^f_{23})^2+(\delta^f_{13})^2-\delta^f_{12}\delta^f_{23}-\delta^f_{23}\delta^f_{13}-\delta^f_{13}\delta^f_{12}
\label{eq.Q}
	\end{eqnarray}
Differing from the quasi-democratic pattern from $S_3$ symmetry with more complicated parameterization \cite{Sogami1998PTP,Fukuura1999PRD,Fritzsch2017CPC}. Here, $I_{\Delta}^f$ is inspired by hierarchal eigenvalues with only non-diagonal real perturbations.  We will see that the similar close-to-flat structure can be generalized to unify up- and down-type fermion Yukawa interactions and finally unify the quark and lepton into a universal form.

From eq. (\ref{eq.Uckm}), only the combination of ${{\bf {F}}_L^u}({\bf {F}}_L^d)^\dag$ contributes to CKM. A general and convenient choice is to parameterize ${\bf F}^u_L$ by diagonal phases,
\begin{eqnarray}
{\bf {F}}^u_L={\rm diag}(1,e^{i\lambda^u_1},e^{i\lambda^u_2}),~~~
{\bf F}^d_L=1
\label{eq.FLRpara}
\end{eqnarray}
 This form can always be realized by choosing an appropriate quark gauge basis. 

The unitary matrix ${\bf F}_{L,R}^q$ can be explained as a transformation between weak gauge eigenstates and Yukawa interaction eigenstates. Defining Yukawa eigenstates as
\begin{eqnarray}
d_{L,R}^{(Y)}={\bf F}^d_{L,R} d_{L,R},~~~
u_{L,R}^{(Y)}={\bf F}^u_{L,R} u_{L,R}
\label{eq.defineYukawaPhase}
\end{eqnarray}
the Yukawa term can be rewritten as
	\begin{eqnarray}
		-\mathcal{L}_Y^q= { y}^d\bar{Q}_L^{(Y)} \left(\begin{array}{c}0 \\ \frac{v_0}{\sqrt{2}}{\bf I}^d_\Delta\end{array}\right)  d_R^{(Y)}
			+ { y}^u \bar{Q}_L^{(Y)}\left(\begin{array}{c} \frac{v_0}{\sqrt{2}}{\bf I}_\Delta^u\\ 0\end{array}\right) u_R^{(Y)}+H.c.
	\label{eq.derYukawaPhase}
	\end{eqnarray}
with the left-handed doublet $Q_L^{(Y)}\equiv (u_L^{(Y)}, d_L^{(Y)})^T$. 

The same flavor structure has been indicated to be suitable for the lepton sector with extended normal order Dirac neutrino mass.
After absorbing the family breaking effect into the flavor independent Higgs vacuum $\langle H\rangle=\frac{v_0}{\sqrt{2}}{\bf I}_\Delta^f$,
 the Yukawa terms can be expressed in an elegant form with family-universal Yukawa couplings,
\begin{eqnarray}
	-\mathcal{L}_Y=y^u \bar{Q}_L^{(Y)}\tilde{H}u_R^{(Y)}
		+y^d\bar{Q}_L^{(Y)}Hd_R^{(Y)}
		+y^\nu \bar{L}_L^{(Y)}\tilde{H}\nu_R^{(Y)}
		+y^e\bar{L}_L^{(Y)}He_R^{(Y)}
		+h.c.
	\label{eq.FamilyYukLag}
\end{eqnarray}
The flavor structure includes the following minimal parameterization: 3+3 perturbations for up- and down-type quark (lepton) masses, 2 phases in ${\bf F}_L^u$ (${\bf F}_L^\nu$) provided to CKM (PMNS), and 2 family-universal Yukawa couplings for absolute quark (lepton) masses. 
This parameterization provides a general structure independent of the characteristics from CKM or PMNS patterns.  More importantly, the validity of the minimal flavor structure has been checked by reproduction of all quark and lepton mass (initializing the lightest neutrino mass), CKM and PMNS data \cite{Zhang2021arXiv} (see Appendix A for details).
\section{Unified Yukawa Coupling for Up- and Down-type Fermions}\label{sec.quarkleptonYuk}

In this section, we unify up- and down-type quark Yukawa interactions into a universal form and then generalize it to the lepton sector.   

\begin{figure}
	\centering  
	\includegraphics[height=0.16 \textheight]{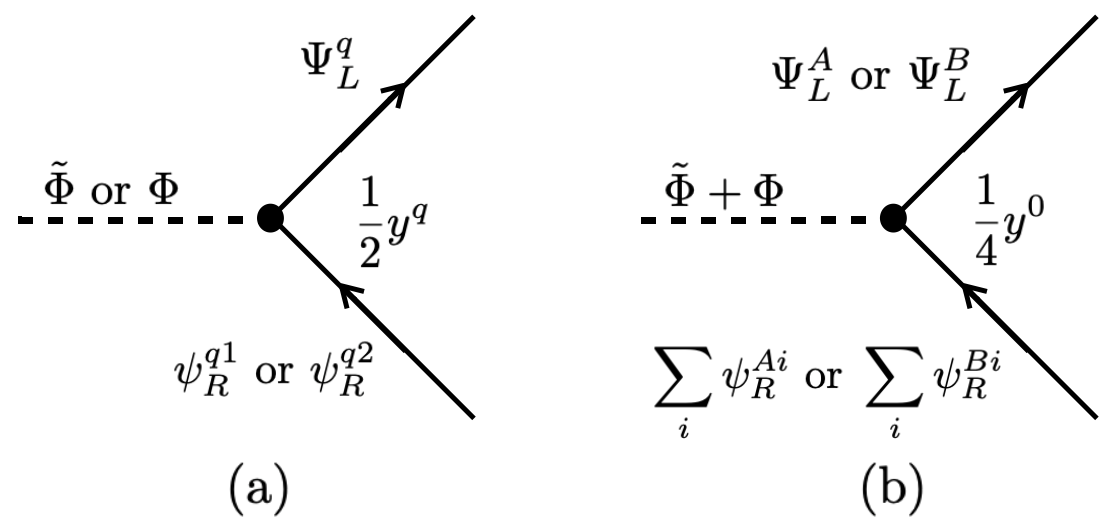}
	\caption{Universal Yukawa interaction: (a) for up- and down-type fermions; (b) for quarks and leptons} 
	\label{fig.UniversalQuarkYukawa}	
\end{figure}

Assuming two quark fields $\psi^{q1}$ and $\psi^{q2}$ in a new Yukawa eigenstate (for the sake of convenience, the superscript $^{(Y)}$ has been neglected), their left-handed components consist of an $SU(2)_L$ doublet $\Psi_L^q=(\psi^{q1}_L,~ \psi^{q2}_L)^T$, and the right-handed  $\psi^{q1}_R$ and $\psi^{q2}_R$ remain singlets.
A unified quark Yukawa interaction with an $SU(2)_L$ doublet scalar $\Phi$ can be written as
			\begin{eqnarray}
				-\mathcal{L}_Y=\frac{1}{2}y^q\bar{\Psi}_L^q(\Phi+\tilde{\Phi})(\psi^{q1}_R+\psi^{q2}_R)
				\label{eq.quarkYukawa1}
			\end{eqnarray}
			or, in matrix form,
			\begin{eqnarray}
				-\mathcal{L}_Y=\frac{1}{2}y^q\bar{\Psi}_L^q\Array{cc}{\tilde{\Phi} & \Phi}\Array{cc}{1&1 \\ 1& 1}\Array{c}{\psi^{q1}_R\\ \psi^{q2}_R}
				\label{eq.quarkYukawaMtx}
			\end{eqnarray}
where $y^q$ is a real coupling. The flat matrix in the above indicates a non-distinguishing Yukawa interaction for one of the right-handed $\psi^{q1}_R$ and $\psi^{q2}_R$ and the scalars $\Phi$ and $\tilde{\Phi}$, as shown in Fig.\ref{fig.UniversalQuarkYukawa}(a).

Inspired by the hierarchal eigenvalues of the close-to-flat matrix, the unified Yukawa interaction can be broken into hierarchal interactions if there exists a perturbation in the flat matrix,
	\begin{eqnarray*}
		\Array{cc}{1&1 \\ 1& 1}\rightarrow\Array{cc}{1&1+\delta^q \\ 1+\delta^q& 1}\equiv {\bf I}_\delta^q
	\end{eqnarray*}
In appropriately chosen right-handed states, ${\bf I}_\delta^q$ can be diagonalized into hierarchal eigenvalues $y^u\gg y^d$
$$\frac{y^q}{2}R_{ud}{\bf I}_\delta^{q}R^T_{ud}=\text{diag}(y^u,y^d)$$
by a real orthogonal rotation,
\begin{eqnarray}
		\Array{c}{\psi^{q1}_{L,R} \\ \psi^{q2}_{L,R}}\rightarrow R_{ud}\Array{c}{\psi^{q1}_{L,R} \\ \psi^{q2}_{L,R}},~~
		R_{ud}\equiv\Array{cc}{\cos\theta_q&-\sin\theta_q \\ \sin\theta_q & \cos\theta_q}
		\label{eq.Rud}
	\end{eqnarray}
where the rotation angle $\theta^q=-\pi/4$.
Rotated quarks are just the SM up- and down-type quarks (in Yukawa eigenstates),
	\begin{eqnarray}
		R_{ud}\Array{c}{\psi^{q1}_{L,R} \\ \psi^{q2}_{L,R}}=\Array{c}{u^{(Y)}_{L,R} \\ d^{(Y)}_{L,R}}.
	\end{eqnarray}
The ratio of $y^u/y^d$ is controlled by the perturbation $\delta_{q}$,
			\begin{eqnarray*}
	\frac{y^u}{y^d}=\frac{-\delta^q}{2+\delta^{q}}
			\end{eqnarray*}
Using $y^{u,d}=\frac{v_0}{3\sqrt{2}}\sum m_{u,d}^i$, the value of $\delta_{q}$ is
			$\delta^{q}\simeq-2m^b/m^t\simeq{-0.0483}$.
The rotated scalar thus becomes the SM Higgs doublet and its charge conjugation,
	\begin{eqnarray}
	\Array{cc}{\tilde{\Phi} & \Phi}\xrightarrow{R_{ud}} {{\bf R}_{ud}\Array{cc}{\tilde{\Phi} & \Phi}{\bf R}_{ud}^T=}\Array{cc}{\tilde{H} & H}
	\end{eqnarray}
Thus, the unified quark Yukawa term has been broken into a family-universal form,
			\begin{eqnarray*}
				-\mathcal{L}_Y
				=\bar{Q}^{(Y)}_L \Array{cc}{\tilde{H}&  H}\Array{cc}{y^u & \\ & y^d}\Array{c}{u^{(Y)}_R \\ d^{(Y)}_R}
			\end{eqnarray*}
as shown in the first two terms of eq. (\ref{eq.FamilyYukLag}).
After ${\bf R}_{ud}$ rotation, the above Lagrangian retains its charge symmetry, which implies that the perturbation $\delta^q$ may come from electromagnetic correction when electroweak symmetry broken. The analysis must thus be performed at an energy scale using a consistent set of input data and RGE to evolve the observables in the future.

	Generalizing these discussions to the lepton sector, we can express the quark and lepton Yukawa terms using two universal Yukawa couplings, $y^q$ and $y^l$, as follows:
			\begin{eqnarray}
			-\mathcal{L}_Y=\frac{1}{2}y^q\bar{\Psi}^{q}_L(\Phi+\tilde{\Phi})(\psi^{q1}_R+\psi^{q2}_R)
				+\frac{1}{2}y^l\bar{\Psi}^{l}_L(\Phi+\tilde{\Phi})(\psi^{l1}_R+\psi^{l2}_R)
			\label{eq.YukLagQuarkLepton}
			\end{eqnarray}
			Note that the perturbation for the lepton is determined by 
				$\delta_{l}=-\frac{2y^\nu}{y^\nu+y^e}$
				and the rotation angle is $\theta_l=\pi/4$ to yield a lighter neutrino and a heavier charged lepton. 

%%%%%%%%%
\section{Unified Yukawa Coupling for Quarks and Lepton}\label{sec.unifiedYuk}
Next, we apply a similar method to unify the quark and lepton Yukawa interactions in eq. (\ref{eq.YukLagQuarkLepton}) into a universal form. 
Considering four fermions $\psi^{Ai}$ and $\psi^{Bi}$ for $i=1,2$,  left-handed components are arranged into two $SU(2)_L$ doublets
$\Psi_{L}^A=(\psi^{A1}_L, \psi^{A2}_L)^T$ and $\Psi_{L}^B=(\psi^{B1}_L, \psi^{B2}_L)^T$,
while the four right-handed components $\psi^{A1}_R,\psi^{A2}_R,\psi^{B1}_R$, and $\psi^{B2}_R$  are singlets. 
The unified Yukawa term can be expressed as
			\begin{eqnarray}
				-\mathcal{L}_Y
				=\frac{y^0}{4}\left(\bar{\Psi}^{A}_L+ \bar{\Psi}^{B}_L\right)(\Phi+\tilde{\Phi})\sum_{i=1,2}\left(\psi^{Ai}_R +\psi^{Bi}_R\right)
				\label{eq.UniversalYukawa1}
			\end{eqnarray}	
or, equivalently in matrix form, as
			\begin{eqnarray}
				-\mathcal{L}_Y
				=\frac{y^0}{4}\Array{cc}{\bar{\Psi}^{A}_L, & \bar{\Psi}^{B}_L}\Array{cc}{1&1\\ 1&1}\Array{c}{\sum_i\psi^{Ai}_R \\ \sum_i\psi^{Bi}_R}(\Phi+\tilde{\Phi})
				\label{eq.UniversalYukawaMtx}
			\end{eqnarray}	
The flat matrix appears again, which indicates an undifferentiated interaction between left-handed $\Psi_L^{A}$ and $\Psi_L^{B}$, right-handed $\sum_i\psi_R^{Ai}$ and $\sum_i\psi_R^{Bi}$ and the scalar $\Phi+\tilde{\Phi}$ (as shown in Fig.\ref{fig.UniversalQuarkYukawa}(b)). 
Now, we assume that some of the fermions experience a correction. A possible origin is the strong interaction quantum effect, which only affect the fermions participating in the strong interaction. This strong interaction splits fermions into strong and non-strong interaction fermions, i.e., quarks and leptons.

The quantum correction as a perturbation can be labeled by $\delta_S$ in the close-to-flat structure, 
	$$\Array{cc}{1 & 1\\ 1 & 1}\rightarrow \Array{cc}{1 & 1+\delta_S \\ 1+\delta_S &1}\equiv {\bf I}_\delta^s$$
 Taking a rotation $\bf {R}_{lq}$,
 \begin{eqnarray}
&\Array{c}{\Psi^{A}_L \\ \Psi^{B}_L}\rightarrow {\bf R}_{lq}\Array{c}{\Psi^{A}_L \\ \Psi^{B}_L}=\Array{c}{\Psi^{l}_L \\ \Psi^{q}_L}
\\
&
\Array{c}{\sum_i\psi^{Ai}_R \\ \sum_i\psi^{Bi}_R}\rightarrow {\bf R}_{lq}\Array{c}{\sum_i\psi^{Ai}_R \\ \sum_i\psi^{Bi}_R}=\Array{c}{\sum_i\psi^{li}_R \\ \sum_i\psi^{qi}_R}
\end{eqnarray}
with the mixing angle $\theta_{lq}=\pi/4$,
 ${\bf I}_\delta^s$ can be diagonalized into 
 $$\frac{y^0}{2}{\bf I}_\delta^s\rightarrow {\frac{y^0}{2}{\bf R}_{lq}{\bf I}_\delta^s{\bf R}^T_{lq}}=\Array{cc}{ y^l & \\ & y^q}$$
The perturbation is determined by $\delta_{S}\simeq-\frac{2m^\tau}{m^t}\simeq {-0.0205}$ to yield two hierarchal eigenvalues $y^l$ and $y^q$ (with  $y^l\ll y^q$).	
After rotation ${\bf R}_{lq}$, the quark and lepton Yukawa interaction in eq.(\ref{eq.YukLagQuarkLepton}) has been derived from the universal Yukawa term in eq. (\ref{eq.UniversalYukawa1}).

%%%%
\section{Summary}\label{sec.summary}
\begin{figure}
	\centering  
	\includegraphics[height=0.2\textheight]{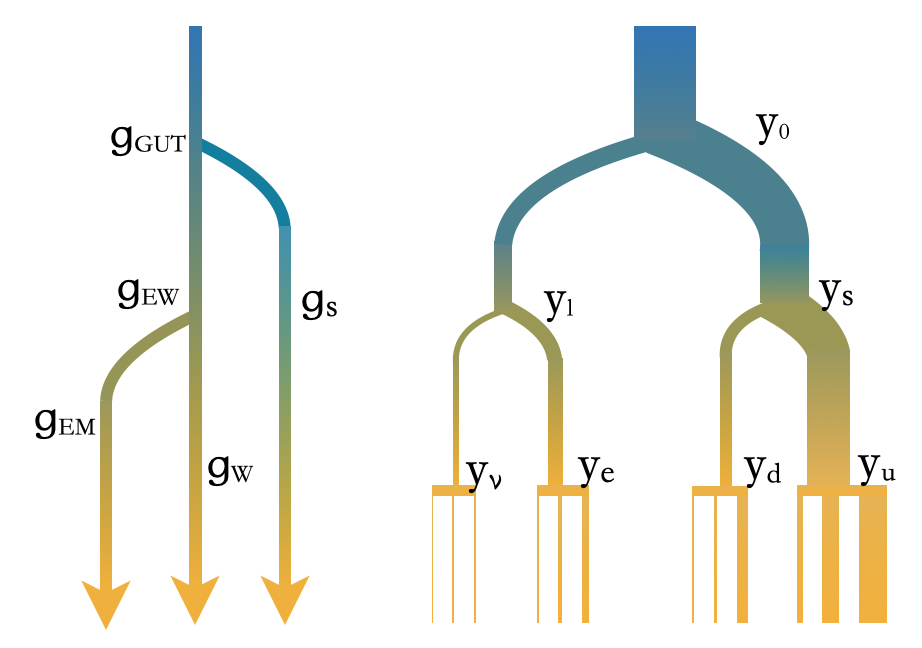}  
	\caption{Gauge interaction unification and Yukawa interaction unification}
	\label{fig.unifyYuk}
\end{figure}

Motivated by the redundancy of the SM Yukawa couplings and similarity of quark and lepton mixings, we unify the SM Yukawa couplings into a universal form for all quarks and leptons. Only one universal coupling exists when fermions are expressed as good Yukawa eigenstates. The ambiguous flavor structure in the SM is involved by mispairing fermions in gauge eigenstates. The entire picture can be summarized as follows: the Yukawa interaction has a flavor-universal structure for all flavors in eq. (\ref{eq.UniversalYukawa1}). When the energy decreases, some fermions participate in the strong interaction, while others do not. The Yukawa interaction for strong fermions obtains a coupling correction, which involves the Yukawa term breaking into a strong quark and a non-strong lepton. Quark and lepton Yukawa couplings correspond to the eigenvalues of a close-to-flat coupling matrix. Due to the small corrections, leptons have a very small Yukawa interaction compared to quarks. At even lower energies, electromagnetic interactions are separated from electroweak interactions. Left- and right-handed quarks (leptons) thus experience different Yukawa couplings. Up- and down-type quark Yukawa couplings are also eigenvalues of a close-to-flat matrix with non-diagonal electromagnetic correction. Thus, in each family, generations can be split due to the presence of a close-to-flat flavor vacuum. This scenario agrees with the unified method of gauge symmetry, as shown in Fig.\ref{fig.unifyYuk}. 
Flat structures are repeated throughout the entire Yukawa interaction unification process, which exhibits a non-distinguishing flavor structure at multiple levels when a proper Yukawa eigenstate is chosen. There is a natural correspondence between the hierarchal masses/couplings and the flat structure. The problem of mass hierarchy is often addressed from the pattern $(m^f_1,m^f_2,m^f_3)=(0,0,3)$, which is another manifestation of a flat structure in mass eigenstates \cite{Weinberg2020PRD}.
Strong experimental support of the existence of a close-to-flat flavor structure comes from the widespread mass hierarchy:
	\begin{itemize}
		\item in generations: $m^f_1\ll m_2^f \ll m_3^f$ $(f=u,d,\nu,e)$ for $y^f\rightarrow y^f_1, y^f_2,y^f_3$;
		\item in up- and down-type fermions: $\sum_{i=1..3} m_\nu^i\ll \sum_{i=1..3} m_e^i$ for $y^l\rightarrow y^\nu, y^e$ and $\sum_{i=1..3} m_d^i\ll \sum_{i=1..3} m_u^i$ for $y^l\rightarrow y^u, y^d$;
		\item in quark-lepton: $\sum_{i=1..3} (m_e^i+m_\nu^i)\ll \sum_{i=1..3} (m_u^i+m_d^i)$ for $y^0\rightarrow y^q, y^l$.
	\end{itemize}
In the framework of a unified Yukawa interaction, the origin of the perturbations at each level remains the most difficult challenge in the future. 

\section*{Acknowledgments}
I thank my collaborator Prof. Rong Li for helpful discussions and suggestions  on the subject. This work is partially supported  by the Fundamental Research Funds for the Central Universities (XJTU).

\begin{appendix}
\section{Minimal Flavor Structure and Fit}
Using bi-unitary decomposition of the fermion mass matrix in eq.(\ref{eq.massmatrixdecomp}),(\ref{eq.Mpara}) and (\ref{eq.FLRpara}), the quark mass matrix can be parameterized into the form
	\begin{eqnarray}
	{\bf M}^d=\frac{v_0}{3\sqrt{2}}y^d{\bf I}_\Delta^d,~~~
	{\bf M}^u=\frac{v_0}{3\sqrt{2}}y^u({\bf F}_L^u)^\dag {\bf I}_\Delta^u
	\end{eqnarray}
The CKM has the form
	\begin{eqnarray}
	{\bf U}_{CKM}= {\bf U}_0^u \text{~diag}\left(1,e^{i\lambda^u_1},e^{i\lambda^u_2}\right)({\bf U}_0^d)^T
	\end{eqnarray}
The lepton mass matrix and PMNS are also expressed as
	\begin{eqnarray}
	&&{\bf M}^e=\frac{v_0}{3\sqrt{2}}y^e{\bf I}_\Delta^e,~~~
	{\bf M}^\nu=\frac{v_0}{3\sqrt{2}}y^\nu({\bf F}_L^\nu)^\dag {\bf I}_\Delta^\nu
	\\
	&&{\bf U}_{PMNS}
		={\bf U}_0^e \text{~diag}\left(1, e^{-i\lambda^\nu_1}, e^{-i\lambda^\nu_2}\right){{\bf U}_0^\nu}^\dag
	\end{eqnarray}
After matching these flavor parameters to quark/lepton masses and CKM and PMNS mixings in \cite{PDG2020}, the following results are obtained \cite{Zhang2021arXiv}:
\begin{table}[htp]
\begin{center}
\begin{tabular}{c|c|c|c}
\hline\hline
&$\delta^d_{ij}$ & $\delta^u_{ij}$ & $ \lambda^u_i$ 
\\
\hline
quark input &$\begin{array}{c}\delta_{12}^d=-0.00723\\
	\delta_{23}^d=-0.0644\\
	\delta_{13}^d=-0.0377\end{array}$ 
   & $\begin{array}{c}\delta_{12}^u=-0.0000453\\
	\delta_{23}^u=-0.0172\\
	\delta_{13}^u=-0.0165\end{array}$
  & $\begin{array}{c}
	\lambda_1^u=-0.00504\\
	\lambda_2^u=0.0851
	\end{array}$
\\
\hline
\&$m^d_i$ & $m^u_i$ & $ CKM$ 
\\
\hline
results &   $\begin{array}{c}m^d=4.750~MeV\\
	m^s=98.96~MeV\\
	m^b=4.176~GeV
	\end{array}$
   & $\begin{array}{c}m^u=2.205~MeV\\
	m^c=1.303~GeV\\
	m^t=173.0~GeV
	\end{array}$
  & $\begin{array}{c}s_{12}=0.2243\\
	s_{23}=0.04141\\
	s_{13}=0.003942\\
	\delta_{CP}=75.07^\circ\end{array}$
\\
\hline
\hline
para. &$\begin{array}{c}
	\delta_{12}^e=-0.00443\\
	\delta_{23}^e=-0.146\\
	\delta_{13}^e=-0.105\end{array}$ 
   & $\begin{array}{c}
   	\delta_{12}^\nu=-0.176\\
	\delta_{23}^\nu=-0.435\\
	\delta_{13}^\nu=-0.073\end{array}$
  & $\begin{array}{c}
	\lambda_1^\nu=-0.0111\\
	\lambda_2^\nu=1.60\end{array}$
\\
\hline
&$m^e_i$ & $m^\nu_i$ & $ PMNS$ 
\\
\hline
results &   $\begin{array}{c}
	m^e=0.5108~MeV\\
	m^\mu=105.55~MeV\\
	m^\tau=1.777~GeV
	\end{array}$
   & $\begin{array}{c}
   	m^\nu_1=0.0001~{eV}\\
	m^\nu_2=0.008640~{eV}\\
	m^\nu_3=0.04996~{eV}
	\end{array}$
  & $\begin{array}{c}s^2_{12}=0.3353\\
	s^2_{23}=0.4393\\
	s^2_{13}=0.02009\\
	\delta_{CP}=1.486\pi\end{array}$
\\
\hline\hline
\end{tabular}
\end{center}
\caption{Matched results for quarks and leptons. $v_0y^f/\sqrt{2}$ is set as the total family mass, and the lightest neutrino mass is set as $0.0001$ eV as an example.}
\label{tab.quarkresult}
\end{table}%	

\end{appendix}
%%%%%%%


\begin{thebibliography}{10}
\bibitem{Buras2005arXiv}
	A. J. Buras, arXiv:hep-ph/0505175.
\bibitem{Raidal2008EPJC}
	M. Raidal, A. van der Schaaf, I. Bigi et al. ,Eur.Phys.J.C \textbf{57} (2008) 13-182 [arXiv: 0801.1826 [hep-ph]].
\bibitem{ZZX2020PR}
	Z. Xing, Phys.Rept. \textbf{854} (2020) 1-147 [arXiv: 1909.09610 [hep-ph]].
\bibitem{Feruglio2015EPJC}
	F. Feruglio, Eur.Phys.J.C \textbf{75}, no.8, 373 (2015) [arXiv:1503.04071 [hep-ph]].
\bibitem{Rodejohann2003PRD}
	W. Rodejohann, Phys.Rev.D \textbf{69} (2004) 033005 [arXiv: hep-ph/0309249 [hep-ph]].
\bibitem{Li2005PRD}
	N. Li, B. Ma, Phys.Rev.D \textbf{71} (2005) 097301
[arXiv: hep-ph/0501226 [hep-ph]].
\bibitem{King2013RPP}
	S.F. King, C. Luhn, Rept.Prog.Phys. \textbf{76} (2013) 056201.
\bibitem{PDG2020}
	P.A. Zyla et al. (Particle Data Group), Prog. Theor. Exp. Phys. 2020 (2020) 8, 083C01.
\bibitem{Zhang2021arXiv}
	Y. Zhang,  arXiv:2102.06830v2 [hep-ph].
\bibitem{Fritzsch2017CPC}
	H. Fritzsch, Z. Xing, D. Zhang, Chin.Phys.C \textbf{41} (2017) 9, 093104 [arXiv: 1705.01391 [hep-ph]].
\bibitem{Fukuura1999PRD}
	K. Fukuura, T. Miura, E. Takasugi, M. Yoshimura, Phys.Rev.D \textbf{61} (2000) 073002 [arXiv:hep-ph/9909415 [hep-ph]].
\bibitem{Sogami1998PTP}
	I.S. Sogami, K. Nishida, H. Tanaka, T. Shinohara, Prog.Theor.Phys. \textbf{99} (1998) 281-292 [arXiv: hep-ph/9801342 [hep-ph]].
\bibitem{Weinberg2020PRD}
	S. Weinberg, Phys.Rev.D 101 (2020) 3, 035020
[arXiv: 2001.06582 [hep-th]].
\end{thebibliography}
\end{document}